\documentclass[10pt]{article}
\usepackage{opex3}
\usepackage{bm}
\usepackage{upgreek}

\begin{document}
\title{Efficient generation and characterization of spectrally factorable biphotons}

\OEauthor{Changchen Chen,$^{1*}$ Cao Bo,$^{2}$ Murphy Yuezhen Niu,$^{1,3}$ Feihu Xu,$^{1}$ Zheshen Zhang,$^{1}$ Jeffrey H. Shapiro,$^{1}$ and Franco N. C. Wong$^{1}$}
\OEaddress{{1}Research Laboratory of Electronics, Massachusetts Institute of Technology, Cambridge, MA 02139, USA\\
{2}Research Institute of Electrical Communication, Tohoku University, Sendai 980-8577, Japan\\
{3}Department of Physics, Massachusetts Institute of Technology, Cambridge, MA 02139, USA}
\OEemail{{*}chencc@mit.edu}

\begin{abstract}
Spectrally unentangled biphotons with high single-spatiotemporal-mode purity are highly desirable for many quantum information processing tasks. We generate biphotons with an inferred heralded-state spectral purity of 99\%, the highest to date without any spectral filtering, by pulsed spontaneous parametric downconversion in a custom-fabricated periodically-poled KTiOPO$_4$ crystal under extended Gaussian phase-matching conditions. To efficiently characterize the joint spectral intensity of the generated biphotons at high spectral resolution, we employ a commercially available dispersion compensation module (DCM) with a dispersion equivalent to 100\,km of standard optical fiber and with an insertion loss of only 2.8\,dB. 
Compared with the typical method of using two temperature-stabilized equal-length fibers that incurs an insertion loss of 20\,dB per fiber, the DCM approach achieves high spectral resolution in a much shorter measurement time. 
Because the dispersion amount and center wavelengths of DCMs can be easily customized, spectral characterization in a wide range of quantum photonic applications should benefit significantly from this technique. 
\end{abstract}


\section{Introduction}

Photonic quantum information processing (QIP) applications, such as quantum computation \cite{politi2009shor,mower2014towards}, boson sampling \cite{white2013boson,spring2013boson,crespi2013integrated,Tillmann2013boson}, and quantum repeaters for long-distance quantum communication networks \cite{gisin2011repeater,Lo2015repeater}, utilize measurement-based protocols that rely on high-visibility interference between individual photons. To successfully carry out these QIP tasks often involves multiple near-perfect interference measurements in parallel and/or in series, which require the interfering photons be indistinguishable in all their characteristic degrees of freedom. Some of these features are easy to control and maintain, such as their polarizations with wave plates and polarizers, and their spatial modes with single-mode fibers or waveguides. Other aspects, such as their spectra and times of arrival at the interference locations, are not as simple to manipulate. 

Consider the widely used method of generating heralded single photons by pulsed spontaneous parametric downconversion (SPDC) in a nonlinear optical crystal. Typical SPDC outputs consist of pairs of spectrally entangled signal and idler photons whose biphoton state has a Schmidt number greater than 1 under Schmidt decomposition \cite{eberly2000decomp, eberly2004Schmidt}. The detection of the idler photon of the spectrally entangled biphoton heralds the presence of the signal photon; however, the heralded signal photon is in a spectrally mixed state \cite{rohde2007spectral}, as dictated by the biphoton's Schmidt mode structure. Therefore, interference of two such heralded photons, even with the same spectral mixture, does not produce high visibility because of their low spectral purity. Recent research in overcoming the low heralded-state spectral purity of SPDC outputs has been focused on methods of generating spectrally factorable biphoton state with an ideal Schmidt number of 1 \cite{grice2001correlations, mosley2008heralded, kuzucu2008joint, gerrits2011generation, yabuno2012fourphoton, jin2013widely, harder2013WGsource, gerrits2015spectral, jin2016generation, kaneda2016heralded, weston2016entangled}. A spectrally factorable biphoton state is frequency uncorrelated, and upon heralding, the signal photon has a well defined and definite spectrum. By adopting the same generation method for all interacting photons, high-visibility interference measurements in photonic QIP tasks are achievable. 

In this work, periodically-poled KTiOPO$_4$ (PPKTP) is the nonlinear crystal of choice for its type-II phase matching with outputs at telecom wavelengths. More importantly, PPKTP can be operated under extended phase matching \cite{grice2001correlations, giovannetti2002generating, giovannetti2002extended} that allows the joint spectral amplitude (JSA) of the biphoton state to be controlled by two independent parameters: the pump spectrum and the PPKTP phase-matching function. To fully optimize the biphoton state's heralded-state spectral purity, the PPKTP crystal is custom fabricated to yield a Gaussian-shaped phase-matching function \cite{dixon2013spectral,Murphy}. In Section 2 we show three-wave mixing measurement results that verify the shape of the custom phase-matching function. The heralded-state spectral purity of the biphoton state is typically inferred from its joint spectral intensity (JSI) using dispersion-based spectrometry. Two different dispersive elements are used: either a single long optical fiber or, for higher resolution, a low-loss dispersion compensation module (DCM), which yielded the results shown in Section 3. At a pump bandwidth of 0.95\,nm we infer a heralded-state spectral purity of 99\%, which is the highest purity to date without the use of filtering. 
We also verify that the SPDC output is spectrally entangled under extended phase-matching conditions at 5.6 nm pump bandwidth in Section 4. 
In Section 5 we summarize our results and discuss potential use of our source and characterization methods in photonic QIP applications. 

\section{Extended Gaussian phase matching of PPKTP} 

A spectrally factorable biphoton state can be formed typically with one of two JSA profiles.  An asymmetric JSA \cite{uren2005asymmetric, mosley2008conditional} has an elongated profile (like a long rectangle)  with the long side oriented along the idler ($\omega_I$) frequency axis. Detecting the idler photon yields a signal ($\omega_S$) photon in a narrow single frequency mode, regardless of the idler frequency measurement. The asymmetric JSA has been demonstrated \cite{mosley2008heralded} showing good results that agree with theoretical  predictions.

The second JSA profile for a factorable biphoton has a circularly symmetric shape and is based on extended phase matching in which the JSA is the product of the pump spectrum at ($\omega_S + \omega_I$) and the crystal phase-matching function that can be approximated by a ($\omega_S - \omega_I$) dependence for a reasonably long crystal with type-II phase matching \cite{giovannetti2002extended, erdmann2000restoring, mosley2008conditional}. The pump and phase-matching parameters are oriented at $+45^{\circ}$ and $-45^{\circ}$ in a plot with $\omega_S$ and $\omega_I$ axes, so that the   controls using $\omega_S \pm \omega_I$ parameters allow for orthogonal and convenient adjustments of the JSA. Our approach follows the extended phase matching technique that has been applied successfully in a number of experiments \cite{kuzucu2008joint, gerrits2011generation, yabuno2012fourphoton, jin2013widely, harder2013WGsource, gerrits2015spectral, kaneda2016heralded, weston2016entangled, kuzucu2005two, eckstein2011highly}. We note that extended phase matching is called the zero group-velocity mismatch regime in ultrafast optics.

In an earlier experimental attempt to generate biphotons with high heralded-state spectral purity by SPDC under extended phase-matching conditions, a purity of 88\% was obtained without any spectral filtering \cite{kuzucu2008joint}. It was pointed out that the standard sinc-function shape of the phase-matching function was the limiting factor, and that a Gaussian phase-matching function should in principle allow the (spectrally unfiltered) purity to reach 100\%. The limitation of the sinc-function shape stems from the presence of its side lobes that break the circular symmetry of the JSA profile. Later experiments that also utilized extended phase matching show similar purity measurement results \cite{yabuno2012fourphoton, jin2013widely, harder2013WGsource, kaneda2016heralded, weston2016entangled, eckstein2011highly}.
Generally, spectral filtering is undesirable partly because of insertion loss due to the filters. In addition, when we consider interference between independent heralded single photons from different sources, it can be difficult to match the spectral shapes of the different filters used to eliminate all the side lobes while passing a Gaussian-shaped main lobe of the sinc-shaped phase-matching function. More problematic is the uncertainty of the relative temporal positions of the heralded photons. Because the unfiltered biphoton state has a Schmidt number greater than 1, the heralded single photon is in a mixed state of two or more eigenstates that may have different temporal locations. Spectral filtering of the biphotons' signals and idlers does not affect their relative temporal locations such that interference measurements which require precise timing of the independent photons are degraded.

A better way than spectral filtering the outputs is to modify the ferroelectric domain structure of the periodically-poled nonlinear crystal to yield a phase-matching function that has a Gaussian shape instead of the usual sinc-function shape. In standard crystals, the domains are aligned in the same direction, whereas in periodically-poled crystals the domains are alternately aligned and oppositely-aligned with a 50:50 duty cycle, each occupying half of a poling period $\Lambda$. Two methods have been proposed and demonstrated for realizing a Gaussian phase-matching function. One is to change the poling period along the length of the crystal, with periods equal to integer multiples of $\Lambda$ and with the duty cycle remaining at 50:50 \cite{Branczyk:11}. The higher order periods yield lower effective nonlinearity that can be used to tailor the phase-matching function to have a Gaussian shape.  The second method keeps the poling period at $\Lambda$ but varies the duty cycle along the length of the crystal which also lowers the effective nonlinearity \cite{dixon2013spectral}. Both methods are expected to achieve similar results in purity measurements of $\sim$97\% without any filtering and greater than 99\% with a mild spectral filter to further suppress the residual side lobes.
Here we adopt the second method of modifying the duty cycle along the length of the crystal using a Gaussian configuration that optimizes, for a given crystal length, the achievable heralded-state spectral purity to $\sim$99\% without filtering~\cite{Murphy}. We note that simulated annealing has also been proposed as an alternative method for producing a Gaussian phase-matching function~\cite{dosseva2016shaping}.

\begin{figure}[tb]
\center
\includegraphics[scale = 0.4]{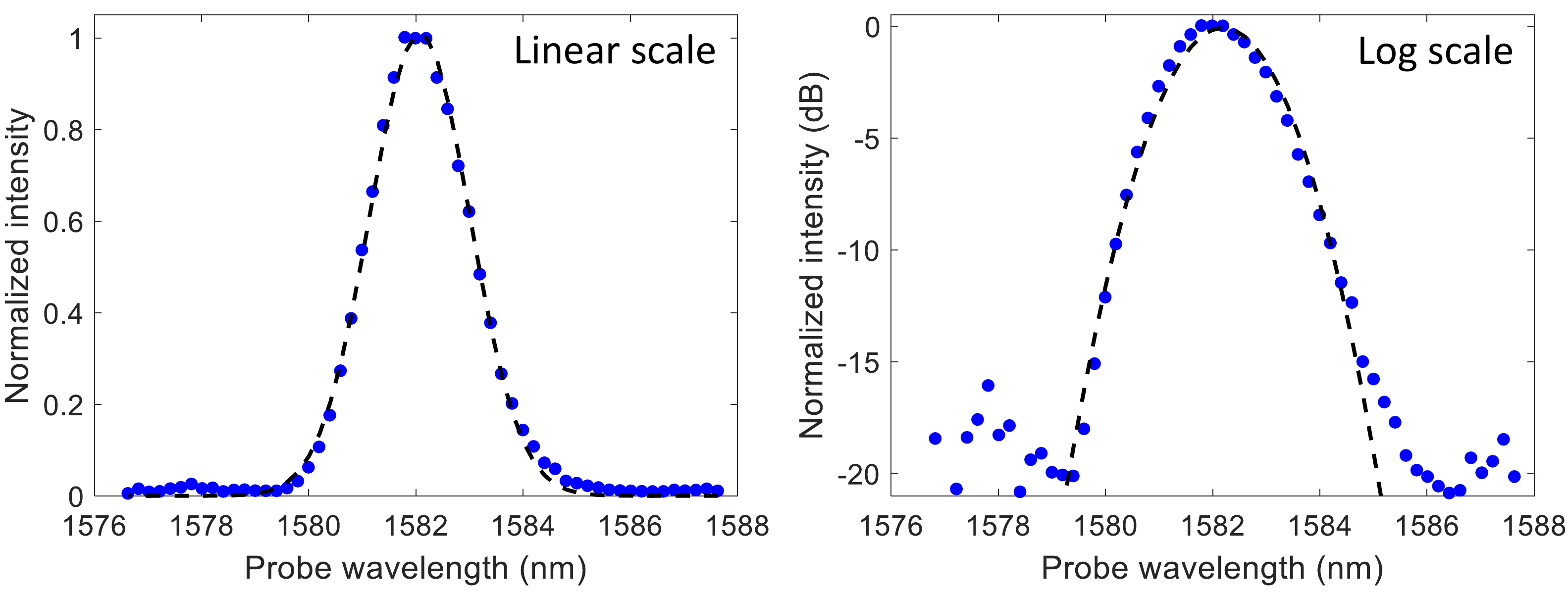}
\caption{Difference-frequency generation outputs as a function of the probe wavelength to confirm Gaussian-shaped phase-matching function based on duty-cycle modulation \cite{dixon2013spectral,Murphy}, in (a) linear scale, and (b) log scale.  DFG outputs are normalized to the peak value.}
\label{DFG}
\vspace{-0.07in}
\end{figure}

Our PPKTP crystal was 18\,mm long with a poling period $\Lambda = 46.1~\upmu$m chosen to yield type-II phase-matched degenerate outputs at 1582\,nm for pumping at 791\,nm. In the custom-fabricated PPKTP we design the crystal with the duty cycle of the poling periods ranging from a ratio of 10:90 to 90:10~\cite{dixon2013spectral,Murphy}. The crystal was anti-reflection coated at both 791 and 1582\,nm wavelengths. Before characterizing the biphoton's joint spectral intensity profile, we first evaluated the phase-matching shape of the custom PPKTP crystal by difference-frequency generation (DFG) measurements. A continuous-wave pump at 791\,nm with $\sim$100\,mW of power was combined with a tunable $\sim$3-mW probe laser centered at 1582\,nm to serve as inputs to the PPKTP crystal. The crystal's temperature was maintained at 26.4$^{\circ}$C so that the peak DFG output occurred at the probe wavelength of 1582\,nm, or twice the pump wavelength. The probe laser wavelength was scanned and the DFG signal was recorded, as shown in Fig.~\ref{DFG} in (a) linear scale and (b) log scale. The measurements show a significant reduction in the usual sinc-function side lobes and a phase-matching function that is Gaussian to a large extent. Figure~\ref{DFG}(b) shows a weak sideband signal at 1578\,nm that we believe is mostly caused by noise in our measurement system. As we will see in Section 3, the side lobe suppression can be inferred to be much better than that shown in Fig.~\ref{DFG}.

\section{Joint spectral intensity measurements}

To quantify the heralded-state spectral purity of the SPDC output, one can measure the signal-idler joint spectral intensity distribution and apply Schmidt decomposition \cite{eberly2000decomp} to calculate the heralded-state purity of the biphoton state, assuming that the state is transform limited.  The most common method of making JSI measurements is to send the pulsed SPDC outputs through two equal-length optical fibers, one for signal and the other for idler, to disperse them after which they are  measured by time-correlated coincidence detection. Fiber dispersion serves to perform frequency-to-time conversion and allows one to easily measure the frequency correlation of photon pairs based on their arrival times \cite{avenhaus2009fiber}. The resolution of the fiber spectrometer depends on the length of the fibers and the timing resolution of the single-photon detectors. A typical setting with 20\,km of standard single-mode fibers (dispersion coefficient of $\sim$17\,ps/nm/km) and superconducting nanowire single-photon detectors (SNSPDs) with a timing resolution of 100\,ps yields a spectral resolution of 0.3\,nm. To obtain a higher spectral resolution requires longer fibers with the drawback that the coincidence detection rate is reduced by the combined insertion loss of the two fibers. To reach 0.12\,nm resolution requires 50\,km of fiber that reduces the coincidence detection rate by a factor of 100. In practice the two individual fibers should be of the same length and temperature stabilized to avoid timing errors caused by their relative length change.

\begin{figure}[tb]
\center
\includegraphics[scale = 0.8]{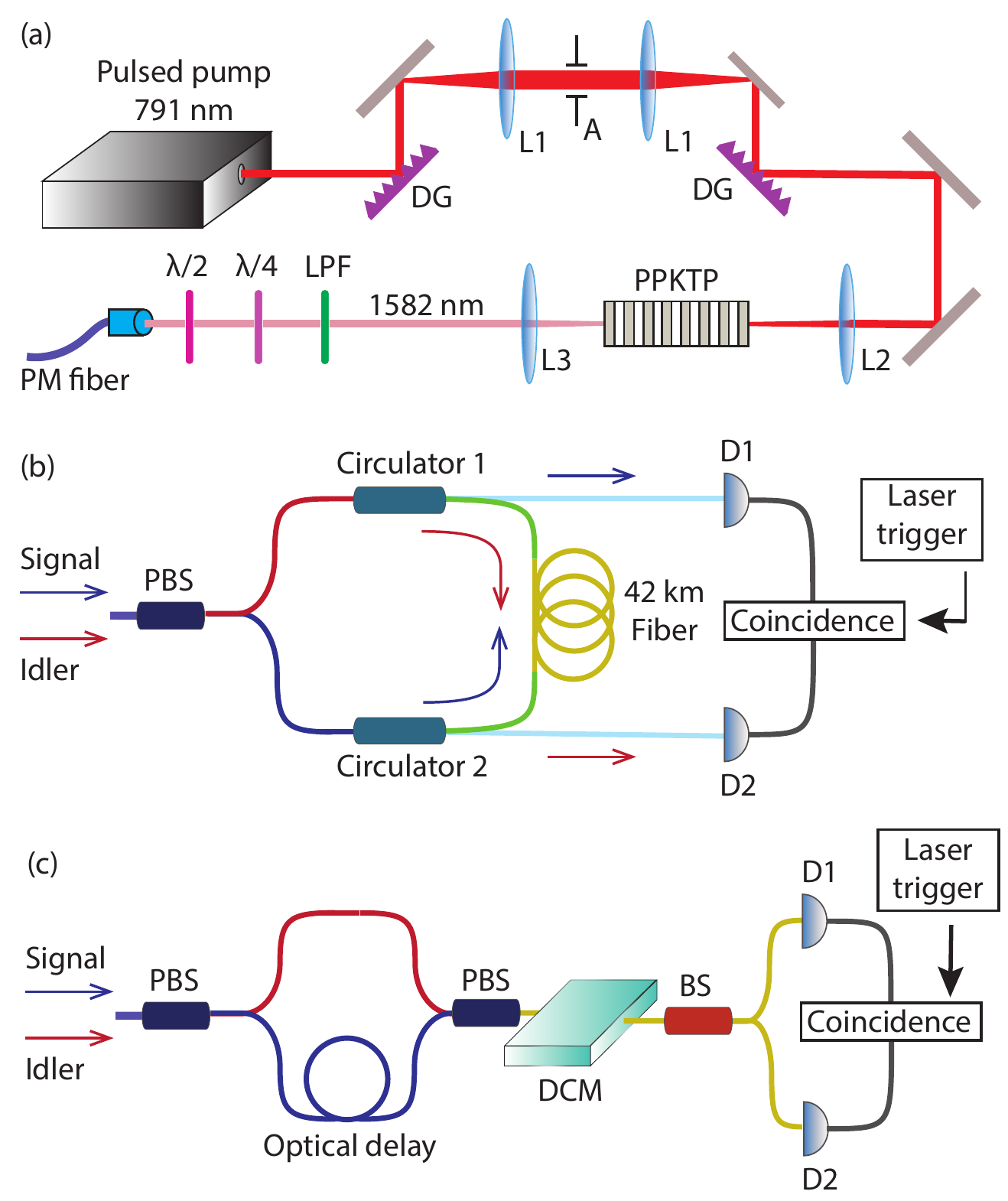}
\caption{Schematics of experiment setups. (a) SPDC photon pairs generation and collection. Pump spectrum is controlled using a pair of diffraction gratings (DGs) in a 4$f$ optical configuration. 
(b) Single-fiber spectrometer with counter-propagating signal and idler and coincidence detection by  SNSPDs D1 and D2. (c) DCM spectrometer providing high resolution JSI measurements. A, rectangular aperture; L1, $f=20$\,cm lens; L2, $f=40$\,cm lens; L3, $f=10$\,cm lens; LPF, long-pass filter; PBS, polarization beam splitter.}
\label{setup}
\vspace{-0.07in}
\end{figure}

Our approach to JSI characterization aims to simplify the measurement setup and to obtain enhanced performance. Two configurations are utilized: one uses a single optical fiber with counter-propagating beams to ensure signal and idler see the same fiber length and dispersion, and the second one employs a commercially available low-loss dispersion compensation module to achieve higher resolution without sacrificing coincidence detection rates. Figure~\ref{setup} shows the experimental setups for  (a) SPDC biphoton generation, (b) single-fiber spectrometer, and (c) DCM spectrometer.

The SPDC pump laser was an 80-MHz mode-locked, $\sim$100 fs Ti-sapphire laser centered at 791\,nm. The optical bandwidth of the pump is one of the parameters for controlling the SPDC biphoton output. To modify the pump bandwidth we implemented a linear spectral filtering system using a pair of identical diffraction gratings in a 4$f$ dispersion-free optical configuration \cite{weiner1988shaping, weiner2000review}.  Two identical lenses L1 with focal length $f=20$\,cm  were placed 2$f$ apart, and the two diffraction gratings were located a distance $f$ from the lenses, as shown in Fig.~\ref{setup}(a). 
The first grating dispersed the collimated pump beam spectrally and the first lens focused the spectrally dispersed components at the focal plane located at a distance \textit{f} from L1. We placed an adjustable rectangular aperture at the focal plane to control the transmission of the parallel spectral components and therefore the bandwidth of the transmitted pump. The second lens and second grating recombined the transmitted spectral components into one collimated beam with the desired bandwidth. With the aperture wide open, the output of the 4$f$ system has the same spectrum as the input's. 

We focused the pump to a beam waist of $\sim$124\,$\upmu$m at the center of the PPKTP crystal that was temperature stabilized at $26.4 \pm 0.1^{\circ}$C to yield wavelength degenerate signal and idler outputs at 1582\,nm. The orthogonally polarized signal and idler were coupled into a single-mode polarization-maintaining (PM) fiber with beam collection optics designed to optimize the symmetric heralding efficiency in single-mode fiber coupling \cite{bennink2010optimal, dixon2014heralding}. As shown in Fig.~\ref{setup}(a) the pump was rejected before fiber coupling by a long-pass filter with a cutoff wavelength of 1300\,nm, and the signal and idler polarizations were aligned with the PM fiber's fast and slow axes using the combination of a quarter-wave plate and a half-wave plate. For a pump bandwidth of 5.6\,nm and at a pump power of 27\,mW, we measured singles of $\sim$95,800/s and $\sim$108,000/s, and $\sim$30,000 coincidences/s, which yields a system efficiency of $\sim$29\%.  

Figure~\ref{setup}(b) shows the single-fiber spectrometer setup for characterizing the joint spectral intensity of the SPDC output from our custom-fabricated PPKTP. We first separated the fiber-coupled signal and idler with a fiber polarization beam splitter (PBS) and sent them through a 42-km SMF28 fiber from its opposite ends. Two fiber circulators were used to provide input/output isolation for both signal and idler light before detection by two WSi SNSPDs with detection efficiencies of $\sim$80\%,  timing jitter of $\sim$200\,ps, and dark count rates of $\sim$400/s. The dark count rates were low enough that no background subtraction adjustment was needed for all measurements. The use of a single fiber ensures that the counter-propagating signal and idler see the same amount of dispersion. The fiber was thermally shielded to reduce variations in length and dispersion due to ambient temperature fluctuations during measurements. The arrival times of signal and idler photons were recorded relative to the trigger pulses from the mode-locked pump laser. Due to fiber dispersion the propagation times of different spectral components propagate at different speeds resulting in different arrival times, therefore allowing the JSI to be reconstructed based on the measured coincidence timing information. We calibrated the fiber propagation time in reference to its zero dispersion point at 1310 nm. 

\begin{figure}[tb]
\center
\includegraphics[scale=0.45]{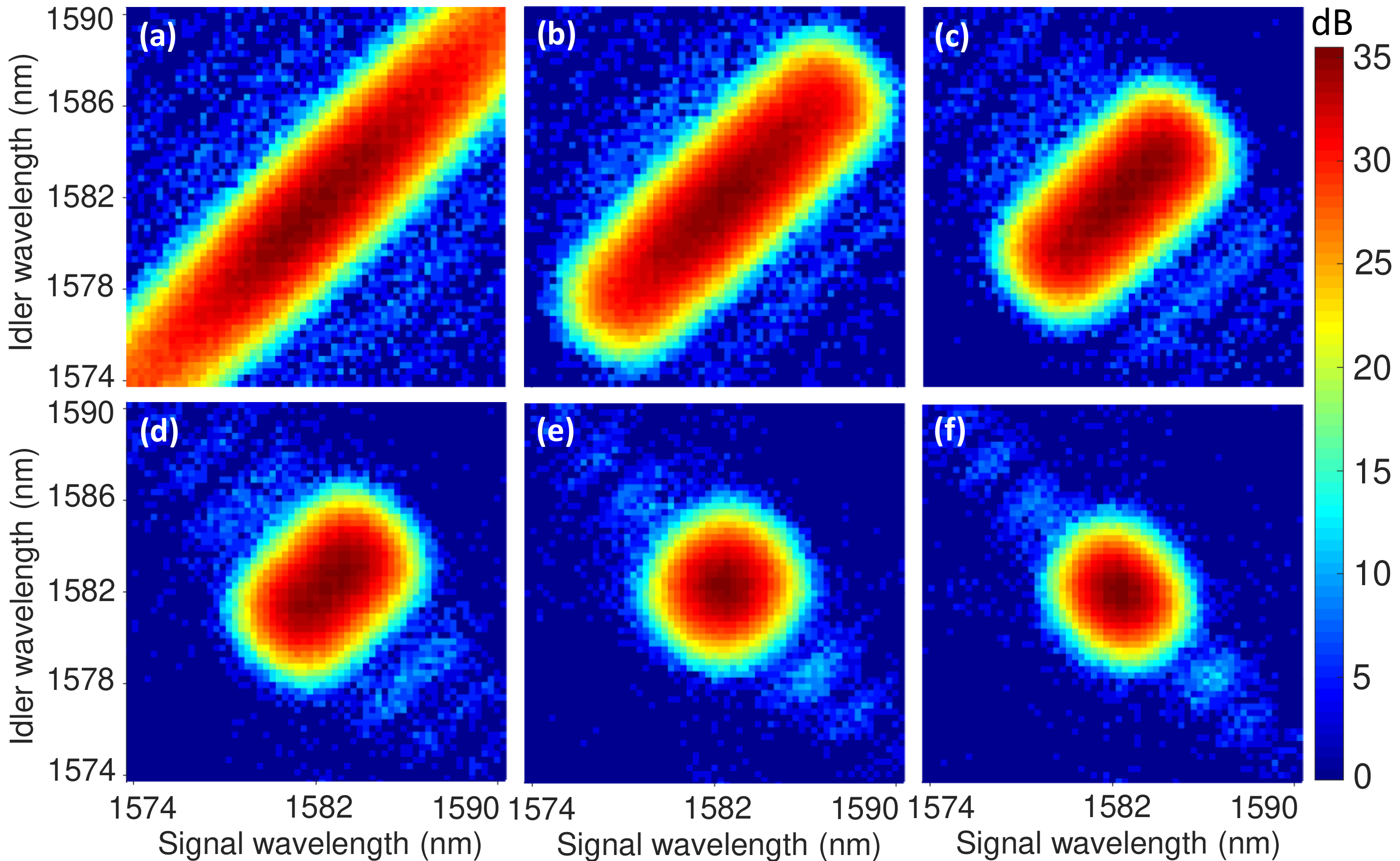}
\caption{JSI measurements from single-fiber spectrometer in log scale. From (a) to (f), the pump bandwidths are 5.6\,nm, 3.6\,nm, 2.4\,nm, 1.6\,nm, 0.95\,nm, and 0.74\,nm, respectively.}
\label{fiberjsi}
\vspace{-0.07in}
\end{figure}

The spectral resolution of our 42-km fiber spectrometer is set by the dispersion imposed by the fiber and the temporal measurement resolution given by the SNSPDs' timing jitter of 200\,ps. The dispersion of the $42$~km fiber was measured to be 0.78~$\pm$~0.07\,ns/nm at 1582\,nm, in line with manufacturer's published data. Therefore our single-fiber spectrometer had a spectral resolution of $\sim$0.25\,nm for both signal and idler. The spectrometer's insertion loss was $\sim$10\,dB per channel that was primarily due to the fiber, suggesting a reduction of 100 fold in the coincidence detection rate when compared to measurements without the spectrometer.  Figure~\ref{fiberjsi} shows the measured log-scale JSI distribution at six different pump bandwidths without the use of any spectral filtering. The wavelength range was 16\,nm centered at 1582\,nm that was set by the pump's mode-locking repetition period of 12.5\,ns and the fiber's dispersion of 0.78\,ns/nm. The peaks of the JSI distributions for all cases are set to center at 1582\,nm that corresponds to the signal and idler degenerate wavelength, thereby setting the correct timing delay we applied relative to the trigger pulses. We then used the dispersion we had measured (0.78\,ns/nm) to map the detected arrival times to the corresponding wavelengths. We only use a linear relation between frequency and time of arrival because any nonlinear adjustment is not significant compared with the relatively large detector timing jitter.

It is known that PPKTP under extended phase matching and pulsed pumping generates coincident-frequency entangled photon pairs \cite{giovannetti2002generating, giovannetti2002extended, kuzucu2005two, kuzucu2008joint}. That is, signal and idler have the same frequencies and their JSI profile is diagonal, as shown in Fig.~\ref{fiberjsi}. We see that the wavelength extent of the signal and idler outputs (along the diagonal) is reduced as the pump bandwidth varies from 5.6\,nm to 0.74\,nm, as expected. The pump spectrum remains approximately Gaussian throughout the change and the pump bandwidths reported are Gaussian fitted measurements. Overall, the JSI measurements match our theoretical expectations very well. Two particular features of Fig.~\ref{fiberjsi} are worth mentioning. The first is that the strongest side lobe of the profiles is at least 24\,dB below the central peak, which corresponds to an 11\,dB suppression of the side lobes of the standard phase-matching sinc function. This is a more sensitive and therefore more accurate measurement of the residual side lobes than the DFG measurements of Fig.~\ref{DFG}(b), suggesting that the Gaussian profile of our custom fabrication design is quite good. The second feature is the nearly symmetric profile of the central lobe in Fig.~\ref{fiberjsi}(e) that is consistent with a factorable biphoton output at the pump bandwidth of 0.95\,nm.

To quantify the spectral correlation of the generated biphoton states in Fig.~\ref{fiberjsi}, we assume that the biphoton state is transform limited so that we can perform a singular value decomposition for continuous variables on the square root of our measurement results to obtain the Schmidt number \cite{eberly2000decomp}. The heralded-state purity is given by the inverse of the Schmidt number \cite{mosley2008conditional}. The Schmidt number and the corresponding purity at different pump bandwidths are listed in Table~1:

\begin{table}[htb]
\vspace{-0.1in}
\center
\caption{Spectral purity of Fig.~\ref{fiberjsi} measurements}
\begin{tabular}{c c c }
\hline
Pump bandwidth (nm) & ~Schmidt number~ & ~~Purity~~ \\
\hline
5.6 & 2.56 & 39\% \\
\hline
3.6 & 1.92 & 52\% \\
\hline
2.4 & 1.41 & 71\% \\
\hline
1.6 & 1.13 & 88\% \\
\hline
0.95 & 1.01 & 99\% \\
\hline
0.74 & 1.03 & 97\% \\
\hline
\end{tabular} 
\vspace{-.1in}
\end{table}

As expected, the purity improves as the JSI profile goes from highly elliptical for broad pump bandwidths to the circularly symmetric case for a pump bandwidth of 0.95\,nm, at which a biphoton state of 1.01 Schmidt number with a corresponding heralded-state purity of 99\% is achieved.  A biphoton state yielding high heralded-state purity such as that shown in Fig.~\ref{fiberjsi}(e) indicates very little frequency correlation between signal and idler, and that such a biphoton state can be ideally suited for generating heralded pure-state single photons. Note that the purity calculation is based on the assumption that the joint distribution of signal and idler amplitudes are transform limited in frequency and time. 
Therefore, our calculated purity can only serve as an upper limit at this point. To verify this assumption, additional measurements of the biphoton state in the time domain, or Hong-Ou-Mandel interferometric measurements between photons from independent sources~\cite{jin2013nonclassical, bruno2014pulsed} are needed.
However, previous measurements have suggested that the biphoton state generated from SPDC is transform limited based on indirect evidence \cite{kuzucu2008joint}. 

To measure the JSI profile more efficiently and with higher resolution, we implemented a different spectrometer based on a commerically available dispersion compensation module instead of a long optical fiber. The DCM uses reflective chirped fiber Bragg gratings to compensate fiber dispersion in long-distance fiber-optic transmission with low insertion loss and low latency. Most relevant to our application is that it has a fixed insertion loss regardless of the amount of its dispersion. Our DCM from Proximion is specified to have a dispersion of 1.88\,ns/nm at 1585\,nm and an insertion loss of 2.8\,dB. For comparison, 100\,km of fiber gives the same amount of dispersion but incurs an insertion loss that is 50$\times$ higher. Furthermore, the DCM is insensitive to ambient temperature fluctuations and hence thermal stabilization is unnecessary. 

We chose to use a single DCM for the spectrometer and intended to send the orthogonally polarized signal and idler through the DCM before separating them with a fiber PBS for subsequent detection. Unfortunately, the DCM had a wavelength dependent polarization mode dispersion of $\sim$0.6\,ps such that the output signal and idler polarizations would be elliptical and no longer orthogonal. In other words, signal and idler cannot be separated based on their polarizations after passing through the DCM. To work around this technical issue, we applied a fixed time delay between signal and idler as identification tags, as shown in the DCM spectrometer configuration of Fig.~\ref{setup}(c). After separating signal and idler with a fiber PBS, we added an optical delay of 11.7\,ns to the signal path before recombining them with a second fiber PBS and sending them through the DCM.  We sent the combined outputs through a 50:50 fiber beam splitter so that about half of the time the signal and idler would be separated and detected by the two WSi SNSPDs. The time separation between signal and idler is much less than the SNSPD recovery time of $\sim$80\,ns and their detection on the same side of the beam splitter output would not be possible. The time-delay scheme incurs effectively a 3\,dB insertion loss but allows us to use a single-DCM implementation. In principle, we could use two DCMs to avoid the 3\,dB penalty, one for signal and the other for idler, as long as the DCMs have identical dispersion \cite{kuo2016spectral, jin2015spectrally}.

\begin{figure}[tb]
\center
\includegraphics[scale=0.45]{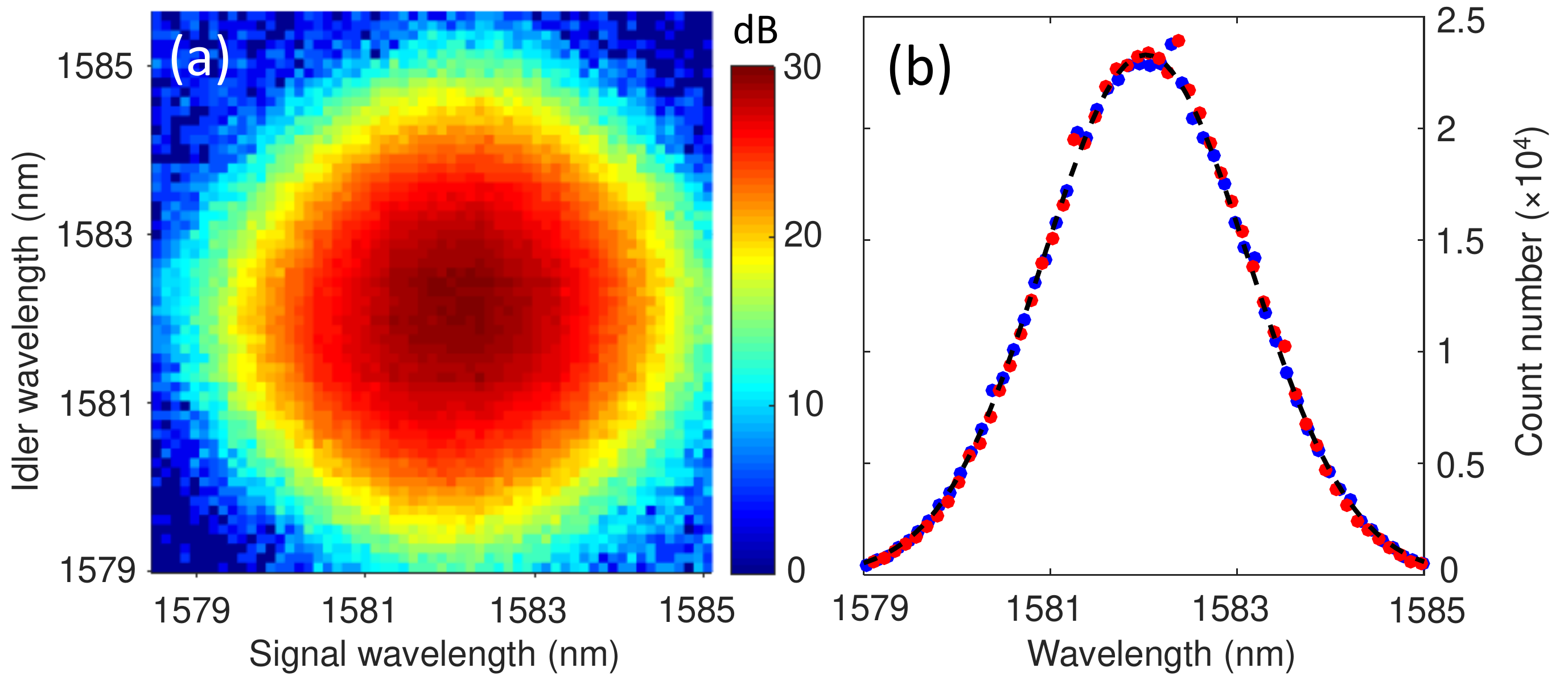}
\caption{JSI and marginal distributions of biphoton state for 0.95\,nm pump bandwidth. (a) High-resolution JSI measurements with 0.11\,nm resolution in log scale. (b) Signal (idler) marginal distribution is shown in blue (red) in linear scale per 0.11-nm bandwidth. Dashed line is Gaussian for comparison.}
\label{DCM}
\end{figure} 

Figure~\ref{DCM} shows the JSI profile and the marginal distributions for the biphoton state for a pump bandwidth of 0.95\,nm using the DCM spectrometer, with a spectral resolution of 0.11\,nm that is more than two times better than that of the single-fiber spectrometer. Because the coincidence measurement interval was limited by the pump repetition period of 12.5\,ns, the wavelength range of Fig.~\ref{DCM} is limited to $\sim$6.6\,nm. The higher resolution measurement of the JSI distrbution of Fig.~\ref{DCM}(a) clearly displays a highly symmetric profile, much like the lower resolution profile of Fig.~\ref{fiberjsi}(e). We note that because the wavelength range is limited to 6.6\,nm, the residual side lobes that are visible in Fig.~\ref{fiberjsi}(e) are mostly out of range. Based on this higher-resolution profile, we obtain a purity estimate of 99.3\% that is slightly higher than that measured with the single-fiber spectrometer of Fig.~\ref{fiberjsi}(e). The slightly higher value is due to the exclusion of the residual side lobes in the smaller wavelength range of the DCM spectrometer, suggesting that the degradation due to the highly suppressed side lobes is very small. The high purity we have achieved without the use of any spectral filtering is entirely the result of the Gaussian phase-matching function of our custom PPKTP cyrstal. We confirm that the marginal distributions shown in Fig.~\ref{DCM}(b) are Gaussian as a result of the custom phase-matching function.

It is useful to recognize the role played by the Gaussian phase-matching function by comparing our results with previous JSI measurements based on PPKTP cyrstals with the standard sinc-function phase matching. Table~2 shows the comparison. Without applying spectral filtering none of these measurements with the standard sinc-shaped phase matching achieved purity higher than 93\%. Weston {\em et al.} obtained an inferred purity close to 100\% using an 8-nm filter applied to the SPDC outputs with a FWHM bandwidth of 15\,nm \cite{weston2016entangled}. That is, near-unity purity has only been achieved  by strong spectral filtering unless the standard phase matching is modified to assume a Gaussian shape as done in our work.

\begin{table}[htb]
\center
\caption{Comparison of spectral purity measurements}
\begin{tabular}{l c r }
\hline
References & ~Spectral filter~ & ~~Purity \\
\hline
Kuzucu {\em et al.} \cite{kuzucu2008joint}\hspace{.1in} & None & 88\% \\
\hline
Gerrits {\em et al.} \cite{gerrits2011generation, gerrits2015spectral} & None & 93\% \\
\hline
Yabuno {\em et al.} \cite{yabuno2012fourphoton} & None & 83\% \\
\hline
Jin {\em et al.} \cite{jin2013widely} & None & 82\% \\
\hline
Harder {\em et al.} \cite{harder2013WGsource} & 8\,nm & 84\% \\
\hline
Kaneda {\em et al.} \cite{kaneda2016heralded} & None & 91\% \\
\hline
Weston {\em et al.} \cite{weston2016entangled} & 8\,nm & $\sim$100\% \\
\hline
This work & None & 99\% \\
\hline
\end{tabular} 
\end{table}

\section{Spectral entanglement}

\begin{figure}[tb]
\center
\includegraphics[scale=0.5]{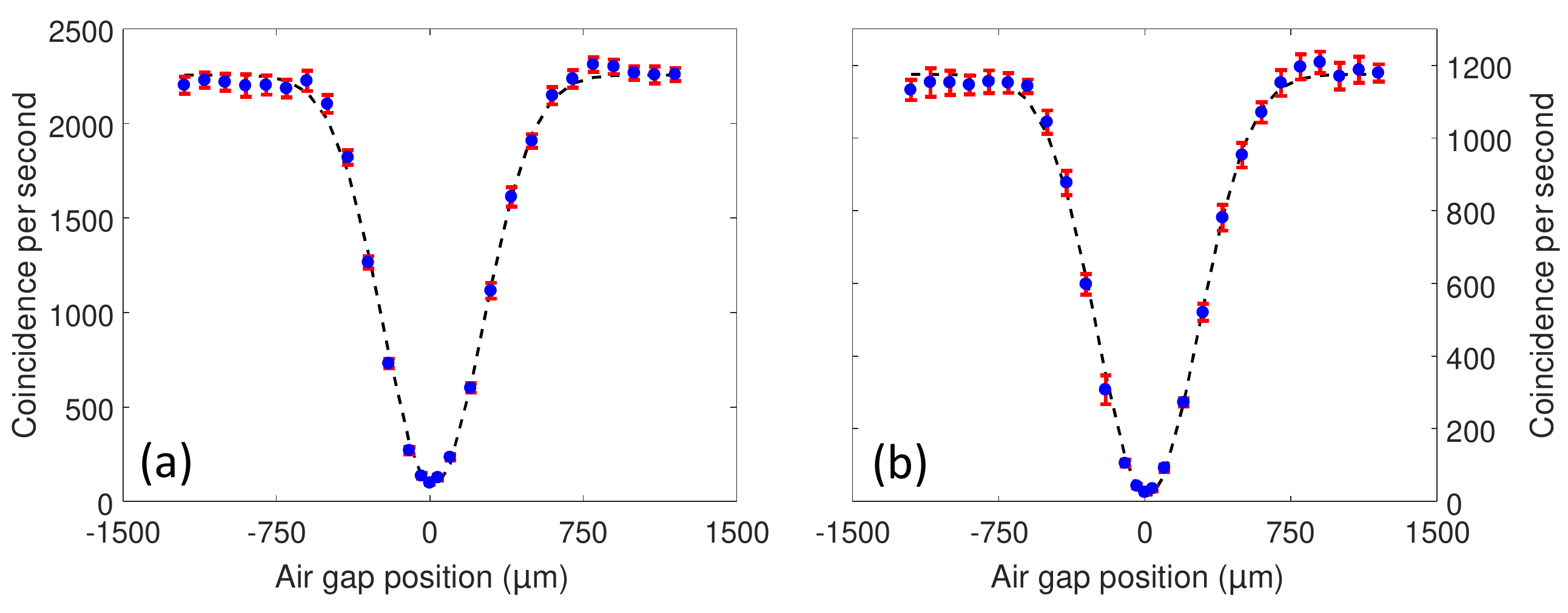}
\caption{HOMI measurements (a) without and (b) with 10-nm bandpass filter, showing Gaussian shape dip with visibility of 92\% and 96\%, respectively. Markers are data points with error bars of one standard deviation and dashed curves are Gaussian fits. The pump bandwidth is 5.6 nm and no background subtraction is applied.}
\label{HOM1}
\end{figure}

Extended phase matching is a technique that generates coincident-frequency entanglement such that it restores signal and idler indistinguishability in pulsed SPDC that can be confirmed with Hong-Ou-Mandel interference (HOMI) measurements. For standard type-II phase matching, the sinc-shaped function leads to an inverted triangular interference dip in HOMI. For Gaussian-shaped phase matching, it is expected that the HOMI should exhibit a Gaussian shape instead \cite{kaneda2016heralded}. We made HOMI measurements with SPDC signal and idler as inputs for a pump bandwidth of 5.6\,nm. To minimize multi-pair emission, the applied pump power was set to 0.25\,nJ per pulse and the mean generated photon pair per pulse was $\sim$0.3\%. After the signal and idler were separated by a fiber PBS, they were sent to interfere at a 50:50 fiber beam splitter with an adjustable air gap delay for one of the input arms. The detected coincidence counts per second versus the air gap delay without and with a 10-nm bandpass filter centered at 1582\,nm are shown in Fig.~\ref{HOM1}.

As expected, the measured data follow the Gaussian shape shown as the dashed curve in Fig.~\ref{HOM1}(a).  The HOMI visibility $V = (N_{\rm max} - N_{\rm min})/(N_{\rm max} + N_{\rm min})$ is found to be 92\% (without background subraction), where $N_{\rm max}$ and $N_{\rm min}$ are the maximum and minimum coincidence counts, respectively. The high visibility implies a high degree of indistinguishability between signal and idler, but it does not reach its asymptotic value of unity visibility that is expected under ideal extended phase-matching conditions \cite{giovannetti2002extended}. By applying a mild filter of 10\,nm bandwidth, the HOMI measurement of Fig.~\ref{HOM1}(b) shows an improved visibility of 96\%, suggesting that removing the residual side lobes are useful in improving the signal-idler indistinguishability. The measured visibility of 96\% is actually quite close to the expected value of 97.8\%, given that we estimate a visibility degradation of 2.2\% owing to imperfect PBS (1\% leakage of the wrong polarization) and a fiber beam splitter with a splitting ratio of 49:51. Additonal degradation could be caused by polarization drifts during measurements. A Gaussian fit of our measured data, as shown in Fig.~\ref{HOM1}, gives a visibility of $94\pm2$\% without filter, and a visibility of $98\pm2$\% with the filter.
The stated uncertainty corresponds to the $95$\% confidence fitting bounds.

From the HOMI measurement of Fig.~\ref{HOM1}(a) we obtain a biphoton coherence time (FWHM) of $1.92\pm 0.06$\,ps, which yields a biphoton coherence bandwidth (FWHM) of $1.92\pm 0.06$\,nm for a Gaussian transform limited time-frequency pair with a time-bandwidth product $\Delta f \Delta \tau = 2 \ln 2 /\pi \approx 0.44$, where $\Delta f$ ($\Delta \tau$) is the frequency bandwidth (time duration). Under extended phase matching, this biphoton coherence bandwidth should also equal the SPDC phase-matching bandwidth that we can obtain from the DFG measurements in Fig.~\ref{DFG}(a) to be $2.2\pm 0.4$\,nm.
The two values are in reasonable agreement and that the difference is possibly due to deviation from ideal extended phase-matching conditions and the relatively large measurement uncertainty of the DFG measurement. In the ideal case \cite{giovannetti2002extended}, the biphoton coherence bandwidth is only a function of the signal-idler frequency difference and hence the phase-matching function, and does not depend on the pump bandwidth. Indeed, one observes in Fig.~\ref{fiberjsi} that the widths of the JSI distributions along the anti-diagonal axis are more or less the same for pump bandwidths ranging from 5.6\,nm to 0.74\,nm.   

At a pump bandwidth of 5.6\,nm  the JSI measurement of Fig.~\ref{fiberjsi}(a) clearly shows that the signal and idler are correlated in frequency. However, the JSI alone is insufficient for one to conclude that they are frequency entangled.
Imagine a joint state that is a low-flux mixture of different types of photon pairs with the following characteristics: the photons in each pair are orthogonally polarized; the photons in each pair type have the same center frequency and the same bandwidth as the HOMI bandwidth (which is much smaller than the JSI bandwidth); and different types of photon pairs have different center frequencies such that the combined frequency coverage spans the bandwidth given by the JSI. Such a mixed state of frequency-unentangled photon pairs would give the same JSI and HOMI results of Fig.~\ref{fiberjsi}(a) and Fig.~\ref{HOM1}(a), respectively.
To show that the pulsed SPDC output state generated under 5.6\,nm pump bandwidth is frequency entangled, we made a signal (idler) field-autocorrelation measurement showing a time duration of $0.36\pm 0.04$\,ps ($0.38\pm 0.04$\,ps) with a corresponding spectral bandwidth of 10.2\,nm (9.69\,nm). Our field-autocorrelation measurement uses a fiber-based Mach-Zehnder interferometer with a single-photon input of either signal or idler and yields the input's spectral bandwidth. The measured spectral widths are much larger than the phase-matching bandwidth, which proves that the joint state could not be a mixture of low-bandwidth photon pairs and confirms that the SPDC output was indeed frequency entangled \cite{lloyd2002quantum}. We note that the marginal distributions for signal and idler from Fig.~\ref{fiberjsi}(a) yield spectral bandwidths of $10.15\pm 0.1$\,nm and $10.27 \pm 0.1$\,nm, respectively, which are in good agreement with the field-autocorrelation measurements. 

To verify the SPDC spectral bandwidth at 0.95\,nm pump bandwidth, we follow the same steps as those taken for confirming the spectral entanglement of the biphoton state when the SPDC was pumped with a pump bandwidth of 5.6\,nm. Field-autocorrelation measurement for signal (idler) at 0.95\,nm pump bandwidth yields a time duration of $1.45\pm0.02$\,ps ($1.41\pm0.01$\,ps), which corresponds to the spectral bandwidth of $2.54\pm0.04$\,nm ($2.61\pm0.02$\,nm). From Fig.~\ref{DCM}(b) we calculate the spectral bandwidth of signal and idler to be $2.62\pm 0.04$\,nm and $2.60\pm 0.03$\,nm, respectively, which are in good agreement with the field-autocorrelation results.


\section{Summary}

To summarize, we have generated and characterized the SPDC biphoton output state under pulsed pumping and extended phase-matching conditions using a custom-fabricated PPKTP crystal that features a Gaussian phase-matching function. Joint spectral intensity measurements confirm that the Gaussian phase-matching function suppresses to a large extent the side lobes that hinder the generation of a circularly symmetric biphoton state. With the side lobes removed, the SPDC biphotons generated under a pump bandwidth of 0.95\,nm can be inferred to have a heralded-state purity of 99\% \textit{without} the use of any spectral filtering in the generation or measurement process, representing the highest purity to date.

Our JSI characterization utilizes two innovative dispersion-based methods to facilitate the measurement procedure. The first is a single-fiber spectrometer that uses a single optical fiber with counter-propagating light, which eliminates the need for two fibers that must be temperature stabilized to maintain equal lengths. The second technique utilizes a commercially available dispersion compensation module with a dispersion amount that can be easily customized and that has low insertion loss of only 2.8\,dB. With dispersion equivalent to 100\,km of fiber, we were able to obtain the JSI with high resolution and a short acquisition time that should be useful for many spectral analysis tasks in quantum information processing. 

Our results demonstrate the unique advantage of SPDC operation under extended phase-matching conditions that allow orthogonal controls of the joint spectral amplitude of the output biphoton state. We were able to modify the crystal's phase-matching function to have a Gaussian shape, and utilize the pump bandwidth to control the spectral property of the SPDC outputs. At a pump bandwidth of 5.6\,nm, we generated strong frequency entanglement between signal and idler that may be useful for photon-efficient quantum communications using high-dimensional frequency encoding \cite{xie2015comb}. When the pump bandwidth is reduced to 0.95\,nm, and assuming the biphoton is transform limited, we obtain signal and idler photons that are ideally suited for generating heralded single photons for many quantum photonic applications that rely on interference between single photons in a single spatiotemporal mode. 
 
\section*{Funding}
Air Force Office of Scientific Research Grant No.~FA9550-14-1-0052.

\end{document}